\documentclass[12pt]{iopart}
\usepackage{amssymb}
\usepackage{graphicx}
\DeclareMathAlphabet{\mathpzc}{OT1}{pzc}{m}{it}

\usepackage{iopams}
\begin{document}

\title[]{Nonlinear coherent state of an exciton in a wide quantum dot}

\author{M. Bagheri Harouni, R. Roknizadeh, M. H. Naderi}

\address{Quantum Optics Group,
Physics Department, University of Isfahan}
\ead{\mailto{m-baghreri@phys.ui.ac.ir},\
\mailto{rokni@sci.ui.ac.ir},\ \mailto{mhnaderi@phys.ui.ac.ir}}
\begin{abstract}
In this paper, we derive the dynamical algebra of a particle
confined in an infinite spherical well by using the $f$-deformed
oscillator approach. We consider an exciton with definite angular
momentum in a wide quantum dot interacting with two laser beams.
We show that under the weak confinement condition, and
quantization of the center-of-mass motion of exciton, the
stationary state of it can be considered as a special kind of
nonlinear coherent states which exhibits the quadrature squeezing.
\end{abstract}

\maketitle

\section{Introduction}
The conventional coherent states of the quantum harmonic
oscillator, defined by Glauber \cite{Glauber} as the right-hand
eigenstates of non-hermitian annihilation operator $\hat{a}
([\hat{a},\hat{a}^{\dag}]=1)$, have found many interesting
applications in different areas of physics such as quantum optics,
condensed matter physics, statistical physics and atomic physics
\cite{klauder}. These states play an important role in the
quantum theory of coherence, are considered as the most classical
ones among the pure quantum states, and laser light can be
supposed as a physical realization of them. Due to the vast
application of these states, there have been many attempts to
generalize them \cite{prelomov}. Among the all generalizations,
nonlinear coherent states (NLCS) \cite{manko} have been paid
attention in recent years because they exhibit nonclassical
features such as quadrature squeezing and sub-poissonian
statistics \cite{hame}. These states are defined as the right-hand
eigenstates of a deformed operator $\hat{A}$
\begin{equation}\label{yek}
  \hat{A}=\hat{a}f(\hat{n})\hspace{2cm}\hat{A}|\alpha,f\rangle=\alpha|\alpha,f\rangle,
\end{equation}
where the deformation function $f(\hat{n})$ is an operator-valued
function of the number operator $\hat n$. From (\ref{yek}) one can
obtain an explicit form of NLCS in the number state representation
\begin{eqnarray}\label{dodo}
|\alpha,f\rangle=\mathcal{N}_f\sum_n\frac{\alpha^n}{\sqrt{n!}f(n)!}|n\rangle,\nonumber \\
\mathcal{N}_f=\left[\sum_n\frac{|\alpha|^{2n}}{[f(n)!]^2n!}\right]^{-\frac{1}{2}}.
\end{eqnarray}
A class of NLCS can be realized physically as the stationary state
of the center-of-mass motion of a laser driven trapped ion
\cite{filho,manko1}. Furthermore, it has been proposed a
theoretical scheme to show the possibility of generating various
families of NLCS \cite{naderi} of the radiation field in a
lossless coherently pumped micromaser within the frame work of the
intensity-dependent Jaynes-Cummings model. \\ \indent Recently,
the influences of the spatial confinement \cite{malek} and the
curvature of physical space \cite{mahdi} on the algebraic
structure of the coherent states of the quantum harmonic oscillator
have been investigated within the frame work of nonlinear coherent
states approach. It has been shown that if a quantum harmonic
oscillator be confined within a small region of order of its
characteristic length \cite{malek} or its physical space to be a
sphere \cite{mahdi}, then it can be regarded as a deformed
oscillator, i.e., an oscillator that its creation and annihilation
operators are deformed operator $\hat{A}$ and $\hat{A}^{\dag}$
given by Eq.(\ref{yek}).\\ \indent On the other hand,
 we can consider nanostructures as systems whose physical properties are related to
 the confinement effects. Thus, we expect that it is possible to realize some natural
deformations in these systems \cite{malek,liu}. In addition, in
nanostructures different kinds of quantum states can be prepared.
One of the most applicable of these states is exciton state.
Exciton is an elementary excitation in semiconductors interacting
with light, electron in conduction band which is bounded to hole
in valance band that can easily move through the sample. In one of
the nano size systems, quantum dot (QD), due to the
confinement in three dimensions, energy bands reduce to quasi energy levels.
Therefore, in order to describe the interaction of QD with light
we can consider it as a few level atom \cite{schmit}. These
Exciton states can be used in quantum information processes. It
has been shown that excitons in coupled QDs are ideal for
preparation of entangled state in solid-state systems \cite{john}.
Entanglement of the exciton states in a single QD or in a QD
molecule has been demonstrated experimentally \cite{chen}.
Entanglement of the coherent states of the excitons in a system of
two coupled QDs has been considered \cite{liu1}. Recently,
coherent exciton states of excitonic nano-crystal-molecules has
been considered \cite{kore}. Theoretical approach for generating
Dick states of excitons in optically driven QD has been proposed
in Ref.\cite{zou}. In a QD, the effects of exciton-phonon
interaction, exciton-impurity interaction and exciton-exciton
interaction play an important role. These effects are the main
sources for the decoherence of exciton states \cite{kdzhu}.
Furthermore, these effects cause the exciton has the spontaneous
recombination or scattered to other exciton modes
\cite{tass,bonda}. \\ \indent In this paper we propose a
theoretical scheme for generating excitonic NLCS. We will show
that under certain conditions the quantized motion of wave packet
of center-of-mass of exciton can be consider as a special kind of
NLCSs. Our scheme is based on the interaction of a quantum dot
with two laser beams. By using the approach considered in
Ref.\cite{filho}, we propose a theoretical
scheme for generation of NLCS of an exciton in a wide QD.\\
\indent In section 2, we consider different confinement regimes in
a QD, and the explicit forms of the creation and
annihilation operators for a particle confined in an infinite well are derived 
by using the deformed quantum oscillator approach. In section 3,
we consider an exciton in a wide QD which interacts with two laser
beams. We shall show that under the weak confinement condition,
the stationary state of the exciton center-of-mass motion can be
considered as a NLCS.

\section{Algebraic approach for a particle in an infinite spherical well}

 \indent
In nanostructures and confined systems, there are three different
confinement regimes. The criteria for this classification is based
on the comparison between excitation Bohr radius and the spatial
dimensions of the system under consideration. In the case of a QD,
these regimes are defined as follows  \cite{han}. \\ \indent We
first introduce three quantities $\Delta E_c$, $\Delta E_v$ and
$V_{exc}$ which, respectively denote: the electron energy due to
the confinement, the hole energy due to the confinement and
Coulomb energy between correlated electron-hole (exciton).\\ 1)
$V_{exc}>\Delta E_c-\Delta E_v$: In this case, the exciton
energy is much greater than the confinement energies of electron
and hole. If we show the system size by $L$ and the exciton Bohr
radius by $a$, then in this regime $L> a$. This regime
corresponds to the weak confinement (in some literature the weak
confinement is characterized by the situation in which the
electron and the hole are not in the same matter, for example,
hole be in QD and excited electron in host matter. In this paper,
by the weak confinement regime we mean $L> a$ and the
excitations in the same matter). In this regime due to the
confinement, the center-of-mass motion of the exciton is quantized
and the confinement do not affect electron and hole separately.
Hence, the
confinement affect the exciton motion as a whole \cite{andrea}.\\
2) $V_{exc}<\Delta E_c\;, \Delta E_v$: This regime, in contrast to
the previous one, is associated with the cases where $L< a$. In
this regime the exciton is completely localized, and the
confinement affects both the electron and the hole independently
and their states become quantized in conduction and
valance bands. This regime is called strong confinement.\\
3) $\Delta E_c> V_{exc}\;,\Delta E_v$: This condition is
equivalent to the situation $a_c< a< a_v$, where $a_c$ and
$a_v$ are, respectively, the Bohr radii of electron and hole.
Here, due to the different effective masses of electron and hole,
the hole which has heavier effective mass is localized and
 the electron
motion will be quantized. This regime is called intermediate confinement.\\
\indent In the first case (weak confinement), in a wide QD, an
exciton can move due to its center-of-mass momentum, and because of 
the presence of the barriers, its center-of-mass motion is
quantized. Therefore, it moves as a whole between energy levels of
an infinite well. We consider a wide spherical QD whose energy
levels are equivalent to the energy levels of a spherical well
\begin{equation}\label{enene}
  E_{nl}=\frac{\hbar^2}{2M}\frac{\alpha_{nl}^2}{R^2},
\end{equation}
where $\alpha_{nl}$ is the n'th zero of the first kind Bessel
function of order $l$, $j_l(x)$. In this energy spectrum according
to the azimuthal symmetry around $z$ axis,
 we have a
degenerate spectrum. As mentioned before, in the weak confinement
regime, the Coulomb potential plays an essential role and its
spectrum is given by
\begin{equation}\label{}
  E_k^b=\frac{\mu
  e^4}{2\hbar^2\varepsilon^2}\frac{1}{k^2}\hspace{0.5cm},\hspace{0.5cm}\mu=\frac{m_em_h}{m_e+m_h},
\end{equation}
where superscript $b$ shows binding energy related to the Coulomb
interaction and $\varepsilon$ shows dielectric constant of the
system. As is usual, we interpret the Coulomb part as an exciton
and another degree of freedom (motion between energy levels of the
well) as the exciton center-of-mass motion. Therefore, in a wide
QD an exciton has two different kinds of degrees of freedom: internal
degrees of freedom due to the Coulomb potential and external degrees
of freedom related to the quantum confinement. Here we consider
the lowest exciton state, $1s$ exciton, because this exciton state
has the largest oscillator strength among other exciton state.
Then the energy of the exciton in a wide QD can be written as
\begin{equation}\label{}
  E_{nlm}=E_g-E_1^b+\frac{\hbar^2}{2MR^2}\alpha_{nl}^2,
\end{equation}
where $E_g$ is the energy gap of QD, $E_1^b=E_k^b|_{k=1}$ is the
exciton binding energy, $M=m_e+m_h$ is the total mass of exciton,
and $R$ is the radius of QD. Due to the relation of quantum
numbers $l$ and $m$ with the angular momentum and the selection rules
for optical transitions, we can fix $l$ and $m$ (by choosing a
certain condition), and hence the energy of exciton depends only
on a single quantum number
\begin{equation}\label{}
  E_{n}=E_g-E_1^b+\frac{\hbar^2}{2MR^2}\alpha_{nl}^2.
\end{equation}
Therefore, we can prepare the conditions under which the exciton
center-of-mass motion has a one-dimensional degree of freedom. Due
to the quantization of the exciton center-of-mass motion, we can
describe the exciton motion between the energy levels by the
action of a special kind of ladder operators. In order to find
these
operators we use the $f$-deformed oscillator approach \cite{manko}. \\
\indent As mentioned elsewhere \cite{malek}, if the energy
spectrum of the system is equally spaced, such as harmonic
oscillator, its creation and annihilation operators satisfy the
ordinary Weyl-Heisenberg algebra, otherwise we can interpret them
as the generators of a generalized Weyl-Heisenberg algebra.\\
\indent The energy spectrum of a particle with mass $M$ confined
in an infinite spherical well can be written as (\ref{enene}).
 According to the conservation of angular momentum, we assume that
particle has been prepared with definite angular momentum (for
example by measuring its angular momentum). Then $l$ becomes
completely determined, i.e., in the energy spectrum the number $l$
is a constant. By determining the number $l$ and considering the
rotational symmetry of the system around the $z$ axis, the angular
part of the spectrum becomes completely determined, and the radius
part is described by (\ref{enene}). Now we use a factorization
method and write the Hamiltonian of the center-of-mass motion of
the system as follows
\begin{equation}\label{facorh}
  \hat{H}=\frac{1}{2}(\hat{A}\hat{A}^{\dag}+\hat{A}^{\dag}\hat{A}),
\end{equation}
where $\hat{A}$ and $\hat{A}^{\dag}$ are defined through the
relation (\ref{yek}). Therefore the spectrum of $\hat{H}$, after
straightforward calculation, is obtained as
\begin{equation}\label{energy1}
  E_n=\frac{1}{2}[(n+1)f^2(n+1)+nf^2(n)].
\end{equation}
By comparing (\ref{energy1}) with Eq.(\ref{enene}) we arrive at
the following expression for the corresponding deformation
function $f_1(\hat{n})$
\begin{equation}\label{}
  f_1(n)=\sqrt{\frac{\hbar^2}{MR^2}\frac{(-1)^n}{n}\sum_{i=1}^n(-1)^i\alpha_{i-1 l}^2}.
\end{equation}
Then, the ladder operators associated with the radial motion of a
confined particle in a spherical infinite well is given by
\begin{eqnarray}\label{operator}
&&\hat{A}=\hat{a}\sqrt{\frac{\hbar^2}{MR^2}\frac{(-1)^n}{n}\sum_{i=1}^n(-1)^i\alpha_{i-1
l}^2},\\ \nonumber
&&\hat{A}^{\dag}=\sqrt{\frac{\hbar^2}{MR^2}\frac{(-1)^n}{n}\sum_{i=1}^n(-1)^i\alpha_{i-1
l}^2}\,\,\hat{a}^{\dag}.
\end{eqnarray}
These two deformed operators obey the following commutation
relation
\begin{equation}\label{}
  [\,\hat{A}\,,\,\hat{A}^{\dag}]=-nf_1^2(n)+\frac{\hbar^2}{MR^2}\alpha_{nl}^2.
\end{equation}
As is usual in the $f$-deformation approach, for a particular
limit of the corresponding deformation parameter, the deformed
algebra should be reduced to the conventional oscillator algebra.
However, in this treatment we note that there is no thing in
common between the harmonic oscillator potential and an infinite
spherical well. Only in the limit $R\rightarrow\infty$, the system
reduces to a free particle which has continuous spectrum. \\
\indent As a result, in this section we conclude that the radial
motion of a particle confined in a three-dimensional infinite
spherical well can be interpreted by an $f$-deformed
Weyl-Heisenberg algebra.

\section{Exciton dynamics in QD}

\begin{figure}
\begin{center}
\includegraphics[angle=0,width=.5\textwidth]{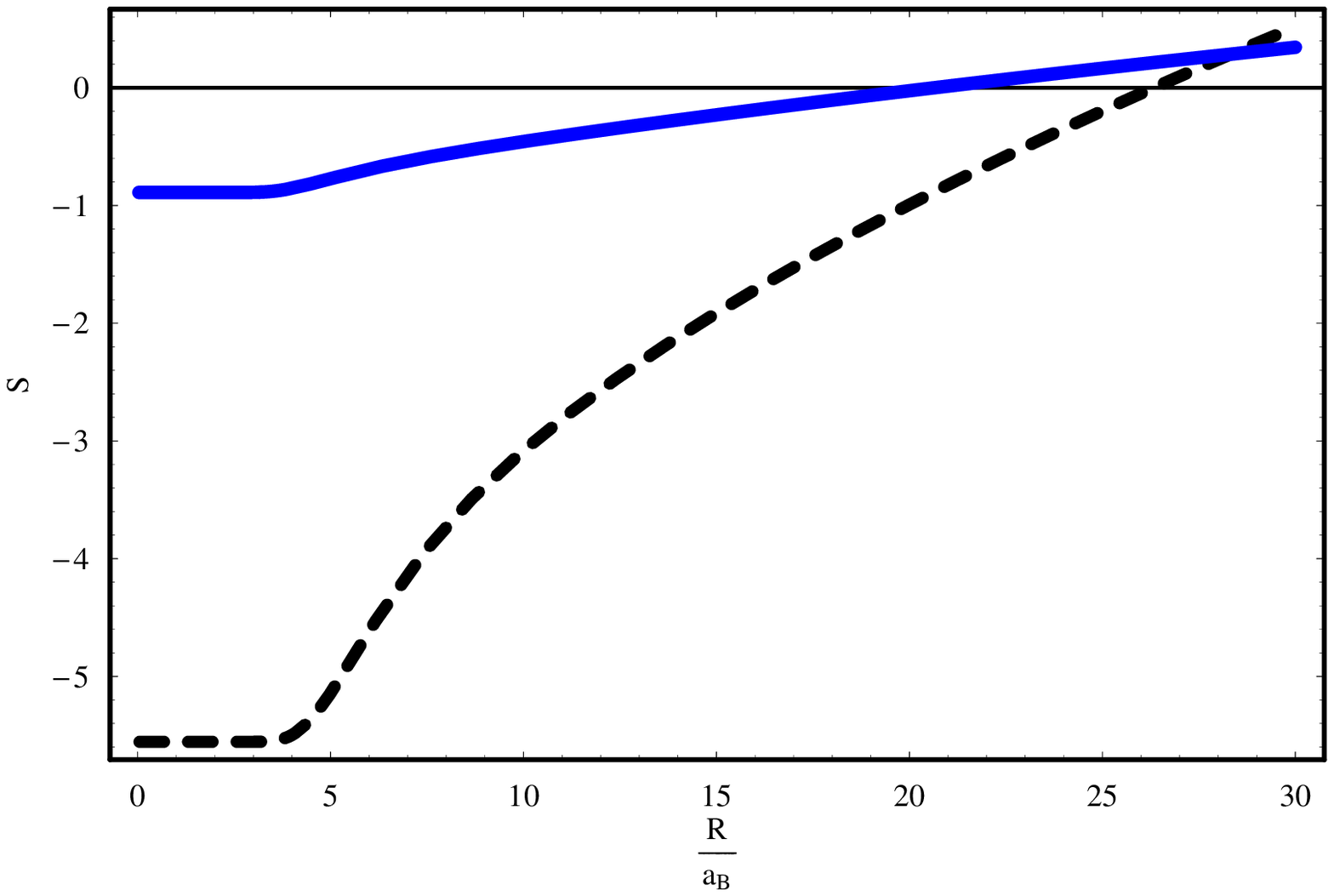}
 \caption{Plots of squeezing versus ration $\frac{r}{a_B}$. Solid line correspond to $\frac{\Omega_0}{\Omega_1}=0.5$,
 dash line correspond to $\frac{\Omega_0}{\Omega_1}=0.2$. In both plots Lamb-Dick parameter is chosen as $\kappa=0.3$.} \label{f1}
\end{center}
\end{figure}

Now we consider the formation of an exciton and its dynamics in a
wide QD during the exciton lifetime. As mentioned before, in this
situation the center-of-mass motion of the exciton is quantized.
The exciton is created during the interaction of a QD with light,
and because of the angular momentum conservation, the exciton has
a well-defined angular momentum. The exciton is a quasiparticle
composed of an electron and a hole and thus the exciton spin state
can be in a singlet state or a triplet state. According to the
optical transition selection rules, the triplet state is optically
inactive and is called dark exciton \cite{dark}. By adding spin
and angular momentum of
 absorbed photons, the angular momentum of the exciton state can be determined. Hence, the
exciton behaves like a particle in a spherical well with the
definite angular momentum. According to the previous section, the
center-of-mass motion of the exciton in the QD and the barriers of
QD can be described by an oscillator-like Hamiltonian expressed in
terms of the $f$-deformed annihilation and creation operators
given by Eq.(\ref{operator})
\begin{equation}\label{doo}
  H_{well}=\frac{1}{2}(\hat{A}\hat{A}^{\dag}+\hat{A}^{\dag}\hat{A}),
\end{equation}
where we interpret the operator $\hat{A}$ $(\hat{A}^{\dag})$ as
the operator whose action causes the transition of exciton
center-of-mass motion to a lower (an upper) energy state. In fact
the Hamiltonian (\ref{doo}) is related to the external degree of
freedom of exciton. On the other hand, one can imagine QD as a
two-level system with the ground state $|g\rangle$ and the excited
state $|e\rangle$ (associated with the presence of exciton). Thus,
for the internal degree of freedom we can consider the following
Hamiltonian
\begin{equation}\label{}
  H_{ex}=\hbar\omega_{ex}\hat{S}_{22},
\end{equation}
where $\hat{S}_{22}=|e\rangle\langle e|-|g\rangle\langle g|$ and
$\hbar\omega_{ex}=E_g-E_1^b$ is the exciton energy.\\
\indent We consider a single exciton of frequency
$\omega_{ex}$ confined in a wide QD interacting with two laser
fields, respectively, tuned to the internal degree of freedom of
the frequency $\omega_{ex}$ and to the non-equal spaced energy
levels of the infinite well. It is necessary that the second laser
has special conditions, because it should give rise to the
transitions between energy levels whose frequencies depend on
intensity. The interacting system can be described by the
Hamiltonian
\begin{equation}\label{}
  \hat{H}=\hat{H}_0+\hat{H}_{int},
\end{equation}
where $\hat{H}_0=\hat{H}_{well}+\hat{H}_{ex}$ and
\begin{equation}\label{hamilton}
  H_{int}=g[E_0e^{-i(k_0\hat{x}-\omega_{ex}t)}+E_1e^{-i(k_1\hat{x}-(\omega_{ex}-\omega_{\overline{n}})t)}
  ]\hat{S}_{12}+H.c.,
\end{equation}
in which $g$ is the coupling constant, $k_0$ and $k_1$ are the
wave vectors of the laser fields, $\hat{S}_{12}=|g\rangle\langle
e|$ is the exciton annihilation operator, and
$\omega_{\overline{n}}$ is the frequency of exciton transition
between
 energy levels of QD due to the spatial confinement. Here, we consider transition
 between specific side-band levels hence, we show the frequency transition with definite
 dependence to $n$. We show this by a c-number quantity $\overline{n}$.\\ \indent The
exciton has a finite lifetime that in systems with small
dimension, is increased \cite{sugawar}. The interaction with
phonons is the main reason of damping of the exciton \cite{jacak}.
Phonons in bulk matter have a continuous spectrum while in a
confined system such as QD their spectrum becomes discrete. Hence
in a QD, the resonant interaction between the exciton and phonons
decreases and in this system the exciton lifetime will increase.
Therefore during the lifetime of an exciton, its dynamics is under
influence of a bath reservoir, and its damping play an important
role. We assume that during the presence of the exciton in QD, it
interacts with the reservoir and hence we can consider its steady
state. We consider an exciton in dark state.
 Experimental preparation methods of such exciton has been described in \cite{dark}.
 In this situation lifetime of exciton will increase and
 exciton has not spontaneously recombination radiation. However, its interaction with
 phonons causes a finite lifetime for it.\\
\indent The operator of the center-of-mass motion position
$\hat{x}$ of the exciton in a spherical QD may be defined as
\begin{equation}\label{}
\hat{x}=\frac{\kappa}{k_{ex}}(\hat{A}+\hat{A}^{\dag}),
\end{equation}
where $\kappa$ being a parameter similar to the Lamb-Dick
parameter in ion trapped systems and is defined as the ratio of QD
radius to the wavelength of the driving laser (because of the
spatial confinement of exciton, its wave function width is
determined by the barriers of QD), and we assume $k_0\simeq
k_1\simeq k_{ex}$ ($k_{ex}$ is the wavevector of the exciton). The
operators $\hat{A}$ and $\hat{A}^{\dag}$ are defined in
Eq.(\ref{operator}).
  The
 interaction Hamiltonian (\ref{hamilton}) can be written as
\begin{equation}\label{}
  H_{int}=\hbar e^{i\omega_{ex}t}\Omega_1\left[\frac{\Omega_0}{\Omega_1}+e^{-i\omega_{\overline{n}}t}\right]e^{i\kappa
  (\hat{A}+\hat{A}^{\dag})}\hat{S}_{12}+H.c.,
\end{equation}
where $\Omega_0=\frac{gE_0}{\hbar}$ and
$\Omega_1=\frac{gE_1}{\hbar}$ are the Rabi frequencies of the
lasers, respectively, tuned to the electronic transition of QD
(internal degree of freedom) and the first center-of-mass motion
transition of exciton. Since the external degree of freedom is
definite, then $\omega_{\overline{n}}$ depends on a special value
of $n$ such that it can be consider as a c-number quantity. The frequency
$\omega_{\overline{n}}$ is depend on the number of quanta for each
transition and hence the laser tuned to the center-of-mass motion
must be so strong that causes transition. This allows us to treat
the interaction of the confined exciton in a wide QD with two
lasers separately, by using a nonlinear Jaynes-Cummings
Hamiltonian \cite{nonlinear} for each coupling. The interaction
Hamiltonian in the interaction picture can be written as
\begin{equation}\label{se}
  H_{I}=\hbar\Omega_1\hat{S}_{12}\left[\frac{\Omega_0}{\Omega_1}+e^{i\omega_{\overline{n}}t}\right]
  \exp[i\kappa(e^{-i\omega_{\hat{n}}t}\hat{A}+\hat{A}^{\dag}e^{i\omega_{\hat{n}}t})]+H.c.,
\end{equation}
where
$\omega_{\hat{n}}=\frac{1}{2\hbar}[(\hat{n}+2)f_1(\hat{n}+2)-\hat{n}f_1(\hat{n})]$.
By using the vibrational rotating wave approximation \cite{filho},
 applying the disentangling formula introduced in \cite{feyn},
and using the fact that in the present case the Lamb-Dick
parameter is small, the interaction Hamiltonian (\ref{se}) is
simplified to
\begin{equation}\label{hamilti}
  H_{I}^{(1)}=\hbar\Omega_1\hat{S}_{12}\left[F_0(\hat{n},\kappa)\frac{\Omega_0}{\Omega_1}+i\kappa
  F_1(\hat{n},\kappa)\hat{a}\right]+H.c.,
\end{equation}
where the function $F_i(\hat{n},\kappa)\;(i=0,1)$ is defined by
\begin{eqnarray}\label{frfrd}
F_i(\hat{n},\kappa)&=&e^{-\frac{\kappa^2}{2}((n+1+i)f_1^2(n+1+i)-(n+i)f_1^2(n+i))}\times\\
\nonumber
&&\sum_{l=0}^n\frac{\left(i\kappa\right)^{2l}}{l!(l+i)!}\frac{f_1(\hat{n})f_1(\hat{n}+i)}{[f_1(\hat{n}-l)!]^2
}(\hat{a}^{\dag})^l\hat{a}^l.
\end{eqnarray}
 It should be
noted that this function in the limit $f_1(\hat{n})\rightarrow 1$
(which is equivalent to the harmonic confinement) is proportional
to the associated Laguerre polynomials
\begin{equation}\label{}
  F_i(\hat{n},\kappa)|_{f_1(\hat{n})\rightarrow 1}=\frac{e^{-\frac{\kappa^2}{2}}}
  {\hat{n}+i}L_{\hat{n}}^i\left(\kappa^2\right).
\end{equation}
Now we write the function $F_i(\hat n,\kappa)$ (\ref{frfrd})
\begin{equation}\label{fdf}
F_i(\hat{n},\kappa)=\frac{e^{-\frac{\kappa^2}{2}((n+1+i)f_1^2(n+1+i)-(n+i)f_1^2(n+i))}}{\hat{n}+i}
f_1(\hat{n})!f_1(\hat{n}+1)!
L_{f,\hat{n}}^i\left(\kappa^2\right),
\end{equation}
where the function $L_{f,\hat{n}}^i(x)$ is defined as
\begin{equation}\label{}
L_{f,\hat{n}}^i(x)=\sum_{l=0}^n\frac{1}{[f_1(\hat{n}-l)!]^2}\frac{(\hat{n}+i)!}{(\hat{n}-l)!l!(l+i)!}(-x)^l.
\end{equation}
This function is similar to the associated Laguerre function.\\
\indent The time evolution of the system under consideration is
characterized by the master equation
\begin{equation}\label{master}
  \frac{d\hat{\rho}}{dt}=-\frac{i}{\hbar}[\hat{H}_{I}^{(1)},\hat{\rho}]+\mathfrak{L}\hat{\rho},
\end{equation}
where $\mathfrak{L}\hat{\rho}$ defines damping of the system due
to the different kinds of interactions which lead to annihilation
of exciton. We assume a bosonic reservoir that causes damping of
exciton system. Due to the properties of dark exciton, the rate of
spontaneous recombination and hence spontaneous emission is
decrease. On the other hand, interactions of exciton-phonon and
exciton-impurities cause the exciton to be damped. In fact in low
temperatures it is possible to ignore the phonon effects and by
assuming a pure system we neglect the impurity effects. Hence we
can write
\begin{equation}\label{}
  \mathfrak{L}\hat{\rho}=\frac{\Gamma}{2}(2\hat{b}\hat{\rho}\hat{b}^{\dag}-\hat{b}^{\dag}\hat{b}\hat{\rho}-\hat{\rho}
  \hat{b}^{\dag}\hat{b}),
\end{equation}
where $\Gamma$ is the energy relaxation rate, $\hat{b}$ and
$\hat{b}^{\dag}$ are the annihilation and creation operators of
the reservoir. Due to the confinement and dark state properties,
spontaneous recombination of exciton decreases and hence the
lifetime of exciton becomes so long that we can consider the
stationary solution of Eq.(\ref{master}). We assume a finite
lifetime for exciton, and during this time we neglect damping
effects. The stationary solution of the master equation
(\ref{master}) in the time scales of our interest is
\begin{equation}\label{}
  \hat{\rho}=|e\rangle|\psi\rangle\langle\psi|\langle e|,
\end{equation}
where $|e\rangle$ is the electronic excited state correspond to
the presence of exciton and $|\psi\rangle$ is the center-of-mass
motion state of the exciton, which can be considered as a
right-hand eigenstate of the deformed operator
$\hat{A}=\frac{F_1(\hat{n},\kappa)}{F_0(\hat{n},\kappa)}\hat{a}$
\begin{equation}\label{defdef}
  \frac{F_1(\hat{n},\kappa)}{F_0(\hat{n},\kappa)}\hat{a}|\psi\rangle=\frac{i\Omega_0}{\Omega_1
  \kappa}|\psi\rangle.
\end{equation}
According to Eq.(\ref{fdf}) the corresponding deformation function
reads as
\begin{eqnarray}\label{fdef}
f(\hat{n})&=&\frac{F_1(\hat{n}-1,\kappa)}{F_0(\hat{n}-1,\kappa)}\\
\nonumber &=&
\frac{f_1(\hat{n})L_{f,\hat{n}-1}^1(\kappa^2)}{nL_{f,\hat{n}-1}^0(\kappa^2)}e^{-\frac{\kappa^2}{2}
\left((n+1)f_1^2(n+1)-(n-1)f_1^2(n-1)\right)}.
\end{eqnarray}
Hence, we can express the state $|\psi\rangle$ in the Fock space
representation as
\begin{equation}\label{}
|\psi\rangle=\mathcal{N}_f\sum_n\frac{\chi^n}{\sqrt{n!}f(n)!}|n\rangle,
\end{equation}
where $\chi=\frac{i\Omega_0}{\kappa\Omega_1}$. According to the
definition (\ref{dodo}), it is evident that the state
$|\psi\rangle$ can be regarded as a special kind of NLCS.
 As is seen from equation (\ref{defdef}), the
eigenvalues of the deformed operator $\hat{A}$ depends on some
physical parameters such
as the Rabi frequencies, the parameter $\kappa$ and radius of QD. \\
\indent As is clear from equation (\ref{fdef}), the deformation
function $f(\hat{n})$ depends on the quantum number $\hat{n}$ and
physical parameters such as QD radius and $\kappa$ which
characterizes the confinement regime. In the limit
$f_1(\hat{n})\rightarrow 1$, (harmonic confinement), which
corresponds, for example, to a QD in lens shape \cite{wojs}, the
function $L_{f,\hat{n}}^i$ reduces to the ordinary associated
Laguerre polynomials, its argument tends to $\kappa^2$ and
therefore, the deformation function (\ref{fdef}) takes the
following form
\begin{equation}\label{}
  f(\hat{n})=e^{-\kappa^2}L_{\hat{n}}^1(\kappa^2)[(\hat{n}+1)L_{\hat{n}}^0(\kappa^2)]^{-1}.
\end{equation}
This is the deformation function that appears in the
center-of-mass motion of a trapped ion confined in a harmonic trap \cite{filho}.\\
\indent In order to investigate the nonclassical behavior of the
NLCS $|\psi\rangle$ we consider the quadrature squeezing of the
center-of-mass motion. For this purpose, we define the deformed
quadratures operators as follows
\begin{equation}\label{}
  \hat{X}_1=\frac{1}{2}(\hat{A}e^{i\phi}+\hat{A}^{\dag}e^{-i\phi}),\hspace{1cm}
  \hat{X}_2=\frac{1}{2i}(\hat{A}e^{i\phi}-\hat{A}^{\dag}e^{-i\phi}).
\end{equation}
In the limiting case $f(\hat{n})\rightarrow 1$, these two
operators reduce to the conventional (non-deformed) quadrature
operators \cite{scully}. The commutation relation of $\hat{X}_1$ and
$\hat{X}_2$ is
\begin{equation}\label{}
  [\hat{X}_1,\hat{X}_2]=\frac{i}{2}[(\hat{n}+1)f^2(\hat{n}+1)-\hat{n}f^2(\hat{n})].
\end{equation}
The variances
$\langle(\Delta\hat{X}_i)^2\rangle\equiv\langle\hat{X}_i^2\rangle-\langle\hat{X}_i\rangle^2
(i=1,2) $ satisfy the uncertainty relation
\begin{equation}\label{}
\langle(\Delta\hat{X}_1)^2\rangle\langle(\Delta\hat{X}_2)^2\rangle\geq\frac{1}{16}
(\langle(\hat{n}+1)f^2(\hat{n}+1)-\hat{n}f^2(\hat{n})\rangle)
\end{equation}
A quantum state is said to be squeezed when one of the quadratures
components $\hat{X}_1$ and $\hat{X}_2$ satisfies the relation
\begin{equation}\label{}
\langle(\Delta\hat{X}_i)^2\rangle<\frac{1}{4}\langle(\hat{n}+1)f^2(\hat{n}+1)-\hat{n}f^2(\hat{n})\rangle
\hspace{1cm} i=1\;or\;2
\end{equation}
The degree of squeezing can be measured by the squeezing parameter
$s_i (i=1,2)$ defined by
\begin{equation}\label{}
  s_i=4\langle(\Delta\hat{X}_i)^2\rangle-\frac{1}{4}\langle(\hat{n}+1)f^2(\hat{n}+1)-\hat{n}f^2(\hat{n})\rangle.
\end{equation}
Then the condition for squeezing in the quadrature component can
be simply written as $s_i<0$. In Fig.(\ref{f1}) we plot the
squeezing parameter $s_1$ versus the parameter $\frac{R}{a_B}$
defined as the ratio of the QD radius to the Bohr radius of
exciton for two different values of ratio
$\frac{\Omega_1}{\Omega_0}$. As is clear from Fig.(\ref{f1}) for
small values of the parameter $\frac{R}{a_B}$ the state shows
quadrature squeezing and by increasing this parameter the
quadrature squeezing disappears.

\section{Conclusion}
In this paper, we first considered a particle confined in a
spherical infinite well and we found the explicit forms of its
creation and annihilation operators by using the $f$-deformed
oscillator approach. Then we considered an exciton in a wide QD
interacts with two laser beams. We showed that under the weak
confinement condition, the exciton is influenced as a whole and
its center-of-mass motion will be quantized. Within the framework
of the $f$-deformed oscillator approach, we found that under
certain circumstances of exciton-laser interaction the stationary
state of the exciton center-of-mass is a nonlinear coherent state
which exhibits the quadrature squeezing.

\textbf{Acknowledgment} The authors wish to thank
      the Office of Graduate Studies of the University of Isfahan and
      Iranian Nanotechnology initiative for
      their support.

{}
\newpage

\newpage

\end{document}